\documentclass[runningheads,a4paper]{llncs}
\usepackage[utf8]{inputenc}
\usepackage[T1]{fontenc}
\usepackage{lmodern}
\usepackage{microtype}
\usepackage[pdfborder={0 0 0}]{hyperref}
\usepackage{amsmath}
\usepackage{amssymb}
\usepackage{graphicx}
\usepackage{url}
\usepackage{tikz}

\begin{document}

\mainmatter

\newcommand{\Omicron}{\mathrm{O}}
\let\tmem\emph
\let\mathbbm\mathbb
\newcommand{\tminput}[2]{\trivlist{\item{#1#2}}}
\newcommand{\tmop}[1]{\ensuremath{\operatorname{#1}}}
\newcommand{\tmsession}[3]{{\topsep=.5ex \partopsep=.5ex\ttfamily\footnotesize#3\par\vspace{1ex}}}
\newcommand{\tmtexttt}[1]{{\ttfamily{#1}}}
\newcommand{\tmunfoldedio}[3]{\trivlist{\item{#1#2}\item{#3}}}

\title{Rigorous Multiple-Precision Evaluation of D-Finite Functions in SageMath}
\titlerunning{Multiple-Precision D-Finite Functions in SageMath}
\author{Marc Mezzarobba%
\thanks{Supported in part by ANR grant ANR-14-CE25-0018-01 (FastRelax).}
}
\authorrunning{Mezzarobba}
\institute{
  CNRS, LIP6, Universit\'e Pierre et Marie Curie%
  \thanks{
    Sorbonne Universit\'es, UPMC Univ Paris 06, CNRS, LIP6 UMR 7606,
    4~place Jussieu 75005 Paris.
  },
  Paris, France\\
  \email{marc@mezzarobba.net},\\
  \texttt{http://marc.mezzarobba.net/}
}
\maketitle

\begin{abstract}
  We present a new open source implementation in the SageMath computer algebra
  system of algorithms for the numerical solution of linear ODEs with
  polynomial coefficients. Our code supports regular singular connection
  problems and provides rigorous error bounds.
  \let\thefootnote\relax
  \footnotetext{This article is in the public domain.
  In jurisdictions where this is not possible, any entity is granted the
  perpetual right to use this work for any purpose, without any conditions
  other than those required by law.}
  \begin{tikzpicture}[remember picture,overlay]
    \node [yshift=6cm, xshift=1em, anchor=north] at (current page.south)
    [text width=\textwidth, font=\footnotesize, align=justify] {
    Extended abstract for a talk given at the 5th~International Congress on Mathematical Software (ICMS~2016).
    Accepted for publication in the proceedings, but withdrawn due to a disagreement with Springer about the above public domain notice.
    };
  \end{tikzpicture}
\end{abstract}

\section{Introduction}

Many special functions satisfy differential equations
\begin{equation}
  p_r (x) f^{(r)} (x) + \cdots + p_1 (x) f^{\prime} (x) + p_0 (x) f (x) = 0
  \label{eq:deq}
\end{equation}
whose coefficients~$p_i$ depend polynomially on the variable~$x$. In virtually
all cases, such special functions can be defined by
complementing~(\ref{eq:deq}) either with simple initial values $f (0),
f^{\prime} (0), \ldots$ or with constraints on the asymptotic behavior of~$f
(x)$ as $x$~approaches a singular point. For example, the error function
satisfies
\begin{equation}
  \tmop{erf}^{\prime\prime} (x) + 2 x \tmop{erf}^{\prime} (x) = 0,
  \hspace{2em} \tmop{erf} (0) = 0, \hspace{1em} \tmop{erf}^{\prime} (0) =
  \frac{2}{\sqrt{\pi}}, \label{eq:erf}
\end{equation}
while the modified Bessel function~$K_0$ is the solution of
\begin{equation}
  xK_0^{\prime\prime} (x) + K_0^{\prime} (x) - xy (x) = 0 \hspace{1em}
  \text{s.t.} \hspace{1em} K_0 (x) = - \log (x / 2) - \gamma + \Omicron_{x
  \rightarrow 0} (x) . \label{eq:Bessel}
\end{equation}
This observation has led to the idea of developing algorithms that deal with
these functions in a uniform way, using the ODE~(\ref{eq:deq}) as a data
structure~{\cite{Lanczos1956,BenoitChyzakDarrasseGerholdMezzarobbaSalvy2010}}.

In this context, solutions of linear ODEs with polynomial coefficients are
called {\tmem{D-finite}} (or {\tmem{holonomic}}) functions. These names
originate from combinatorics, where D-finite power series arise naturally as
generating functions~{\cite{Stanley1999,FlajoletSedgewick2009}}. While
classical special functions typically satisfy ODEs of order 2~to~4 with simple
polynomial coefficients, D-finite generating functions are usually given by
more general and sometimes very large equations. Other notable sources of
general D-finite functions include mathematical physics and number theory.
Explicit examples in theoretical works on meromorphic ODEs also tend to have
polynomial coefficients.

Packages for manipulating D-finite power series
symbolically exist for major computer algebra systems, and include in
particular gfun~{\cite{SalvyZimmermann1994}} and the rest of the algolib
bundle for Maple, GeneratingFunctions~{\cite{Mallinger1996}} and the
RISCErgoSum suite for Mathematica, and
\tmtexttt{ore\_algebra}~{\cite{KauersJaroschekJohansson2015}} for
SageMath~{\cite{Sage}}.

We extended \tmtexttt{ore\_algebra} with features dedicated to D-finite
{\tmem{analytic functions}} such as special functions, starting with numerical
evaluation algorithms. The need for such features is most evident in the
context of special functions, but they turn out to be useful in a wide range
of applications, as we will see from the examples in the sequel. We refer to
our implementation as \tmtexttt{oa.analytic}, after the name of the subpackage
of \tmtexttt{ore\_algebra} where most of the code resides. The present paper
introduces the main features of \tmtexttt{oa.analytic}, with example
applications. The reader is invited to consult the package's documentation for
more information. We further refer to Hille~{\cite{Hille1976}} for more
background on complex analytic ODEs, and to Kauers~{\cite{Kauers2013}} for an
overview of the field of D-finite functions in computer algebra. Also note
that \tmtexttt{oa.analytic} is intended to supersede
NumGfun~{\cite{Mezzarobba2010}}, a Maple package with similar features
previously developed by the author\footnote{Many of the examples in this
article are adapted from earlier work on NumGfun~{\cite{Mezzarobba2011}}.}.

Since the main feature of \tmtexttt{oa.analytic} is to compute numerical
values of solutions of differential equations, it is, at its core, a numerical
ODE solver. However, it is limited to linear ODEs with polynomial coefficients, and differs from typical ODE solvers in a number of other respects. First
of all, the results it computes come with {\tmem{rigorous error bounds}},
making it a ``validated'' ODE solver. It works in {\tmem{arbitrary
precision}}, and implements specific algorithms to make evaluations at very
high precisions feasible. Last but not least, it offers extensive support for
connection problems between {\tmem{regular singular points}} of differential
operators. These distinctive features all rely on {\tmem{symbolic-numeric
algorithms}} that make use of specific properties of D-finite functions.

The development of \tmtexttt{oa.analytic} is still very much a work in
progress. At the time of writing, there is no formal release of
\tmtexttt{ore\_algebra} containing the \tmtexttt{analytic} subpackage yet. The
latest development version of \tmtexttt{oa.analytic} can be found at
{\url{http://marc.mezzarobba.net/code/ore_algebra-analytic}}. It
is distributed under the GNU General Public Licence, version~2 or later.
Comments, bug reports and feature requests are welcome.

\section{Differential Operators in \tmtexttt{ore\_algebra}}

In \tmtexttt{ore\_algebra}, differential operators are represented
algebraically as {\tmem{Ore polynomials}}. More specifically, the operators we
are interested in are elements of the ring, denoted $\mathbbm{K} [x] \langle
D_x \rangle$, of polynomials over $\mathbbm{K} [x]$ for some number field
$\mathbbm{K} \subset \mathbbm{C}$ in a variable~$D_x$ subject to the
commutation rule $D_x x = xD_x + 1$ (which expresses the equality $(xf
(x))^{\prime} = xf^{\prime} (x) + f (x)$). Elements of $\mathbbm{K} [x]
\langle D_x \rangle$ act on differentiable functions in a natural way, namely
by $x \cdot f = (\xi \mapsto \xi f (\xi))$ and $D_x \cdot f = f^{\prime}$, and
can thus be identified with differential operators.

The first thing to do in order to use \tmtexttt{ore\_algebra} in a Sage
session is to install it in the Python search path and \tmtexttt{import} it.
The rings $\mathbbm{K} [x]$ and $\mathbbm{K} [x] \langle D_x \rangle$ and
their respective generators $x$~and~$D_x$ can then be created using standard
Sage syntax as follows:

\tmsession{sage}{default}{
  \tminput{sage: }{from ore\_algebra import OreAlgebra}
  \tminput{sage: }{Pols.<x> = PolynomialRing(QQ); Dops.<Dx> =
  OreAlgebra(Pols)}
}

{\noindent}Note that \tmtexttt{ore\_algebra} supports general Ore polynomials,
but basic arithmetic with differential operators in one variable is all we are
going to need here.

After these commands, the Python variable~\tmtexttt{Dx} contains an object of
type ``Ore polynomial'' that satisfies

\tmsession{sage}{default}{\tmunfoldedio{sage: }{Dx*x}{x*Dx + 1}}

{\noindent}Most features of \tmtexttt{ore\_algebra}, including those presented
in this article, are available as methods of Ore polynomial objects. Some
functions of the \tmtexttt{analytic} subpackage can also be called directly in
advanced usage.

\section{Numerical Analytic Continuation}

Let $L = p_r (x) D_x^r + \cdots + p_0 (x) \in \mathbbm{K} [x] \langle D_x
\rangle$ be a differential operator, and assume that the origin is an ordinary
point of~$L$, i.e.\ that $p_r (0) \neq 0$. By the classical Cauchy existence
theorem~\cite{Hille1976}, the equation $L \cdot f = 0$ then admits an $r$\mbox{-}dimensional
space of complex analytic solutions defined in a neighborhood of~$0$ and
characterized by $r$~initial values at~$0$. The
\tmtexttt{numerical\_solution()} method of differential operators computes
values of D-finite functions defined in this way. Perhaps the simplest example
is the computation of $e = \exp (1)$ as the value at~$1$ of the solution of
$f^{\prime} - f = 0$ such that $f (0) = 1$:

\tmsession{sage}{default}{\tmunfoldedio{sage:
}{(Dx-1).numerical\_solution(ini=[1], path=[0,1],
eps=1e-40)}{[2.71828182845904523536028747135266249775725 +/- 7.53e-42]}}

{\noindent}The last argument, \tmtexttt{eps}, provides an indication of the
accuracy the user is interested in. The method outputs a ``mid-rad interval'',
or {\tmem{ball}}, consisting of a multiple-precision center and a
low-precision radius. In normal circumstances, the diameter of this ball will
be of the order of magnitude of \tmtexttt{eps}, but this is not guaranteed to
be the case. In contrast, barring any bug, the mathematical result is
guaranteed to lie within the bounds of the ball.

Evaluations on the complex plane are also supported. Based on~(\ref{eq:erf}),
we may for instance evaluate the error function as follows:
\enlargethispage{\baselineskip}

\tmsession{sage}{default}{\tmunfoldedio{sage: }{(Dx\^{}2 +
2*x*Dx).numerical\_solution([0, 2/sqrt(pi)], [0, i])}{[+/- 2.91e-25] +
[1.6504257587975429 +/- 2.70e-17]*I}}

{\noindent}Allowed evaluation points and initial values include rational and
floating-point numbers, algebraic numbers, and constant symbolic expressions.

The reader may have noticed that the argument used to specify the evaluation
point is called {\tmem{path}}. The main reason why
\tmtexttt{numerical\_solution()} does not simply take an evaluation
{\tmem{point}} is that in general, solutions of differential equations with
singular points are not single-valued analytic functions. What
\tmtexttt{diffop.numerical\_solution(ini, path)} really computes is the
analytic continuation along \tmtexttt{path} of the solution of
\tmtexttt{diffop} defined in the neighborhood of the starting point by the
initial conditions \tmtexttt{ini}. Consider for example the operator $L = xD_x
+ D_x$, obtained by differentiating the inhomogeneous equation $xy^{\prime}
(x) = 1$. A basis of solutions of~$L$ is $\{ 1, \log (x) \}$, corresponding
respectively to the initial values $(y (1), y^{\prime} (1)) = (1, 0)$ and $(y
(1), y^{\prime} (1)) = (0, 1)$. Analytic continuation of the latter to~$x=-1$ along a
path that passes above the singular point $x = 0$ corresponds to the standard
determination of the complex logarithm, while a path that passes below the
origin yields another branch:

\tmsession{sage}{default}{
  \tmunfoldedio{sage: }{(x*Dx\^{}2 + Dx).numerical\_solution([0, 1], [1, i,
  -1])}{[+/- 5.11e-18] + [3.1415926535897932 +/- 4.38e-17]*I}
  \tmunfoldedio{sage: }{(x*Dx\^{}2 + Dx).numerical\_solution([0, 1], [1, -i,
  -1])}{[+/- 5.11e-18] + [-3.1415926535897932 +/- 4.38e-17]*I}
}

More generally, it is possible to compute the matrix that maps a vector
\begin{equation}
  [f (x_0), f^{\prime} (x_0), \tfrac{1}{2} f^{\prime\prime} (x_0), \ldots,
  \tfrac{1}{(r - 1) !} f^{(r - 1)} (x_0)] \label{eq:ini}
\end{equation}
of initial conditions\footnote{Note the inverse factorials in our definition of
initial conditions. This convention also applies to the \tmtexttt{ini}
argument of \tmtexttt{numerical\_solution()}. It is convenient in view of the
generalization to regular singular points discussed in the next section.}
at~$x_0$ to the vector $[f (x_1), f^{\prime} (x_1), \ldots]$ of ``initial''
conditions at some other point~$x_1$ that define the analytic continuation of
the same solution~$f$ along a given path.
Following van der Hoeven~\cite{vdH1999},
we call this matrix the
{\tmem{transition matrix}} along that path.

\tmsession{sage}{default}{
  \tmunfoldedio{sage: }{(x*Dx\^{}2 + Dx).numerical\_transition\_matrix([1, 2],
  1e-10)}{[ 1.0000000000 [0.69314718056 +/-
  6.99e-14]] \\ \relax
  [ \ \ \ \ \ \ \ \ \ \ \ 0 \ [0.50000000000 +/-
  9.1e-15]]}
}

\begin{example}
  Transition matrices along loops that turn around a singular point exactly
  once are called monodromy matrices. The first step of van~Enckevort and van~Straten's
  experimental study of differential equations of Calabi-Yau
  type~{\cite{EnckevortStraten2006}} consists in computing
  high-precision approximations of all monodromy matrices starting from a
  common base point. The \tmtexttt{numerical\_transition\_matrix()} method
  makes this computation straightforward. In the case of the singularity
  at~$1$ of the operator studied in Example~1 from their article, we
  obtain\footnote{In situations such as this one where the singular point is
  regular singular, it is also possible, and typically more efficient, to
  express the monodromy matrix using the primitives described in the next
  section, as suggested by van der Hoeven~{\cite{vdH2001}}.}:
  \tmsession{sage}{default}{
    \tminput{sage: }{%
    diffop = ((x*Dx)\^{}4 - x*(65*(x*Dx)\^{}4+130*(x*Dx)\^{}3 + 105*(x*Dx)\^{}2\\
    \hbox{}\hfill + 40*x*Dx + 6) + 4*x\^{}2*(4*x*Dx+3)*(x*Dx+1)\^{}2*(4*x*Dx+5))}
    \tminput{sage: }{mat = diffop.numerical\_transition\_matrix([1/2,1-i/2,3/2,1+i/2,1/2],\\
    \hbox{}\hfill1e-1000)}
    \tmunfoldedio{sage: }{mat[0,0] \ \ \# 990 digits omitted}{[1.000...000 +/-
    6.45e-998] + [0.733...6314 +/- 8.99e-998]*I}
  }
\end{example}

The analytic continuation algorithm behind these functions is based
on work by Chudnovsky and Chudnovsky~\cite{ChudnovskyChudnovsky1990}
with some later improvements \cite{vdH1999,Mezzarobba2011}. In the
language of numerical methods for ODEs, it is a high-order Taylor series
method with a step size of the order of the local radius of convergence of
solutions. The computation of the transition matrix for each step boils down
to that of sums of power series whose coefficients are generated by linear
recurrence relations. Two algorithms for this task are currently implemented:
a naïve iterative summation and a simple binary splitting method. In both
cases, the entries of a given column are actually computed all at once, by
working with truncated power series in a manner similar to direct-mode
automatic differentiation. Once the terms of the sum have decreased below the
target accuracy, the algorithm of~{\cite{Mezzarobba-numbounds}} is used to get
a rigorous bound on the tail of the series. Both the computation of the
numerical result and that of error bounds rely on Arb~{\cite{Johansson2013}}
for controlling the round-off errors in basic arithmetic operations.

\section{Regular Singular Connection Problems}

Singular connection problems are similar to the analytic continuation problems
of the previous section, except that one or both ends of the analytic
continuation path are singular points of the operator (points where the
leading coefficient vanishes). At a singular point, the Cauchy theorem does
not apply, and solutions of the operator may diverge or display a branch
point.

Our implementation supports the situation where the singularities involved are
{\tmem{regular singular points}}. A singular point~$x_0$ of $L \in \mathbbm{K}
[x] \langle D_x \rangle$ is regular when $L$~admits a full basis of solutions
each of the form
\begin{equation}
  (x - x_0)^{\alpha}  \left( g_0 (x) + g_1 (x) \log (x - x_0) + \cdots + g_p
  (x)  \frac{(\log (x - x_0))^p}{p!} \right) \label{eq:regsingexp}
\end{equation}
where $g_0, \ldots, g_p$ are analytic functions and $\alpha$ is an algebraic number.

A basic example is that of the modified Bessel equation~(\ref{eq:Bessel}),
which possesses a regular singular point at~$x = 0$ and whose standard solutions
are defined by their asymptotic behavior there. Starting
from~(\ref{eq:Bessel}), we can compute $K_0 (1)$ by:

\tmsession{sage}{default}{
  \tminput{sage: }{ini = [-1, log(2)-euler\_gamma]}
  \tmunfoldedio{sage: }{(x*Dx\^{}2 + Dx - x).numerical\_solution(ini, [0, 1],
  1e-10)}{[0.42102443824 +/- 2.11e-12]}
}

{\noindent}This command is to be understood as follows. It is classical
that, when~$x_0$ is a regular singular point of~$L$, a solution of~$L$ is
characterized by the coefficients of the monomials $(x - x_0)^{\nu} \log (x -
x_0)^k / k!$ in its expansion~(\ref{eq:regsingexp}) about~$x_0$, where
$\nu$~ranges over the roots of the so-called indicial polynomial of~$L$
at~$x_0$ and $k$~is less than the multiplicity of the root~$\nu$. The initial
conditions accepted by \tmtexttt{numerical\_solution()} in this case are the
coefficients of these monomials (ordered in a certain way, basically by
asymptotic dominance as $x \rightarrow x_0$). This definition generalizes
the convention~(\ref{eq:ini}) used at ordinary points. We call the
corresponding basis of solutions the {\tmem{canonical local
basis}}\footnote{Note that there is another common choice, the
{\tmem{Frobenius basis}}.} at~$x_0$. A convenience method provides the
distinguished monomials.

\tmsession{sage}{default}{\tmunfoldedio{sage: }{(x*Dx\^{}2 + Dx -
x).local\_basis\_monomials(0)}{ [log(x), 1]}}

{\noindent}In the case of Equation~(\ref{eq:Bessel}), we see from the
asymptotic condition that the coefficient of~$x^0$ is $\log 2 - \gamma$ and
that of $x^0 \log x$ is $- 1$.

In calls to \tmtexttt{numerical\_transition\_matrix()}, both ends
of the path can be regular singular points. The resulting matrix expresses the
change of basis between the respective canonical local bases.

\begin{example}[Face-Centered Cubic Lattices]
  Koutschan~{\cite{Koutschan2013b}} shows that the lattice Green's function $P
  (x)$ of the four-dimensional face-centered cubic lattice is annihilated by the
  operator
  
  \tmsession{sage}{default}{
  \scriptsize
    \tminput{sage:\ }{%
diffop = ((-1+x)*x\^{}3*(2+x)*(3+x)*(6+x)*(8+x)*(4+
3*x)\^{}2*Dx\^{}4 \\
\hbox{}\hspace{8ex}+ 2*x\^{}2*(4+3*x)*(-3456-2304*x+3676*x\^{}2+4920*x\^{}3+
2079*x\^{}4+356*x\^{}5+21*x\^{}6)*Dx\^{}3 \\
\hbox{}\hspace{8ex}+ 6*x*(-5376-5248*x+11080*x\^{}2+25286*x\^{}3+19898*x\^{}4+
7432*x\^{}5+1286*x\^{}6+81*x\^{}7)*Dx\^{}2 \\
\hbox{}\hspace{8ex}+ 12*(-384+224*x+3716*x\^{}2+7633*x\^{}3+6734*x\^{}4+2939*x\^{}5 + 604*x\^{}6+45*x\^{}7)*Dx \\
\hbox{}\hspace{8ex}+ 12*x*(256+632*x+702*x\^{}2+382*x\^{}3+98*x\^{}4+9*x\^{}5))}
  }
  
  {\noindent}Koutschan then evaluates~$P (1)$ in order to obtain the return probability
  of the lattice. Both $x = 0$ and $x = 1$ are regular singular points of the
  operator. Let us examine the possible behaviors of solutions at these two
  points:
  
  \tmsession{sage}{default}{
    \tmunfoldedio{sage: }{diffop.local\_basis\_monomials(0)}{
    [1/6*log(x)\^{}3, 1/2*log(x)\^{}2, log(x), 1]}
    \tmunfoldedio{sage: }{diffop.local\_basis\_monomials(1)}{ [1, (x -
    1)*log(x - 1), x - 1, (x - 1)\^{}2]}
  }
  
  {\noindent}As it turns out, the space of analytic solutions at the origin
  has dimension~1, and all other solutions involve logarithms. It is easy to
  see from the definition of~$P$ that $P (0) = 1$, so that the function we are
  interested in is exactly the last element of the local canonical basis. At
  $x = 1$, there is a three-dimensional subspace of solutions that tend to
  zero, spanned by the last three basis elements, while the first basis
  function has a finite limit. Thus the value of~$P$ there is the upper right
  entry of the transition matrix from $0$~to~$1$, namely:
  
  \tmsession{sage}{default}{\tmunfoldedio{sage:
  }{diffop.numerical\_transition\_matrix([0, 1],
  1e-60)[0,3]}{
 [1.1058437979212047601829954708859 +/- 4.90e-32] + [+/- 4.55e-41]*I
  }}
\end{example}

\begin{example}[Asymptotics of Apéry Numbers]
  In his celebrated proof that $\zeta (3)$ is irrational~\cite{Apery1979}, Apéry introduces
  the two sequences
  \[ a_n = \sum_{k = 0}^n \binom{n}{k}^2  \binom{n + k}{k}^2, \hspace{2em} b_n
     = \sum_{k = 1}^n \left( \frac{a_n}{k^3} - \sum_{m = 1}^k \frac{(- 1)^m
     \binom{n}{k}^2  \binom{n + k}{k}^2}{2 m^3  \binom{n}{m}  \binom{n +
     m}{m}} \right) . \]
  The generating series $a (x) = \sum_{n = 0}^{\infty} a_n x^n$ and $b (x) =
  \sum_{n = 0}^{\infty} b_n x^n$ are solutions of the following operator:
  
  \tmsession{sage}{default}{
    \tminput{sage: }{diffop = (x\^{}2*(x\^{}2-34*x+1)*Dx\^{}4 +
    5*x*(2*x\^{}2-51*x+1)*Dx\^{}3\\
    \hbox{}\hfill + (25*x\^{}2-418*x+4)*Dx\^{}2 + (15*x-117)*Dx + 1)}
  }
  
  This operator has three regular singular points: $0$, $\alpha = (\sqrt{2}
  + 1)^4 \approx 33.971$, and $\alpha^{- 1} = (\sqrt{2} - 1)^4 \approx
  0.029$.
  
  \tmsession{sage}{default}{\tmunfoldedio{sage: }{s =
  diffop.leading\_coefficient().roots(AA, multiplicities=False); s}{ [0,
  0.02943725152285942?, 33.97056274847714?]}}
  
  {\noindent}This implies that the radius of convergence of $a (x)$ is in $\{
  \alpha, \alpha^{- 1}, \infty \}$, so that $a_n$ must be of one of the forms $\alpha^{- n
  + o (n)}$, $\alpha^{n + o (n)}$ and $e^{o (n)}$ as $n \rightarrow
  \infty$. The same holds for~$b_n$. A key step of Apéry's proof is to show
  that $a_n = \alpha^{n + o (n)}$ and $b_n = \alpha^{n + o (n)}$, but $b_n -
  \zeta (3) a_n = \alpha^{- n + o (n)}$.
  
  Analytic solutions at $x = 0$ are determined by the first two coefficients
  of their Taylor expansions. At $x = \alpha^{- 1}$, the canonical local basis takes the form $(f_0,
  f_1, f_2, f_3)$ where $f_0, f_2, f_3$ are analytic and
  $f_1 (x) \sim \sqrt{x - \alpha^{- 1}}$.
  
  \tmsession{sage}{default}{
    \tmunfoldedio{sage: }{diffop.local\_basis\_monomials(0)}{
    [1/2*log(x)\^{}2, log(x), 1, x]}
    \tmunfoldedio{sage: }{diffop.local\_basis\_monomials(s[1])}{ [1, sqrt(x -
    0.02943725152285942?), x - 0.02943725152285942?,\\
    (x - 0.02943725152285942?)\^{}2]}
  }
  
  {\noindent}To prove that $a_n = \alpha^{n + o (n)}$, it is enough to check
  that the coefficient $c_1$ in the decomposition $a (x) = \sum c_i f_i (x)$
  is nonzero.
  
  \tmsession{sage}{default}{\tminput{sage: }{mat =
  diffop.numerical\_transition\_matrix([0, sing[1]], 1e-40)}}
  
  {\noindent}From the initial values $a_0 = 1$, $a_1 = 5$,  we see that $c_1$
  is equal to
  
  \tmsession{sage}{default}{
    \tmunfoldedio{sage: }{mat[1,2] +
    5*mat[1,3]}{[4.546376247522844600239593024915161553303 +/- 9.85e-41]*I}
  }
  
  {\noindent}With a little more work, using the method of singularity
  analysis~{\cite{FlajoletSedgewick2009}}, this is actually enough to prove
  that $a_n = n^{- 3 / 2} \alpha^n  (\lambda + \Omicron (1 / n))$ where
  $\lambda$ is given by
  
  \tmsession{sage}{default}{\tmunfoldedio{sage:
  }{CBF(-i/(2*sqrt(pi)))*sqrt(sing[1])*(mat[1,2] +
  5*mat[1,3])}{[0.220043767112643 +/- 2.06e-16]}}
  
  {\noindent}This result agrees with the closed form $a_n \sim \sqrt{\alpha}
  (2 \pi \sqrt{2})^{- 3 / 2} n^{- 3 / 2} \alpha^n$ from~{\cite{Hirschhorn2012}}.
  
  In the same way, using the initial terms $b_0 = 0$, $b_1 = 6$ of $(b_n)$, we
  can compute the coefficient of~$f_1$ in the decomposition of $b (x)$ and
  check that the ratio of the two values we obtained is~$\zeta (3)$.
  
  \tmsession{sage}{default}{\tmunfoldedio{sage: }{6*mat[1,3]/(mat[1,2] +
  5*mat[1,3])}{[1.2020569031595942853997381615114499907650 +/- 6.08e-41]}}
\end{example}

\paragraph{Acknowledgements.}This work was prompted by discussions with
Fredrik Johansson. I am also indebted to the Sage developers, in particular
Vincent Delecroix, Jeroen Demeyer and Clemens Heuberger, who reviewed or
helped writing a number of Sage patches vital to this project, and to Christoph Lauter for comments on a version of this text.


\begin{thebibliography}{10}

\enlargethispage\baselineskip

\bibitem{Apery1979}
R.~Apéry.
\newblock Irrationalité de $\zeta(2)$ et $\zeta(3)$.
\newblock {\em Astérisque}, 61, page 11–13, 1979.

\bibitem{BenoitChyzakDarrasseGerholdMezzarobbaSalvy2010}
A.~Benoit, F.~Chyzak, A.~Darrasse, S.~Gerhold, M.~Mezzarobba, and B.~Salvy.
\newblock The dynamic dictionary of mathematical functions ({DDMF}).
\newblock In K.~Fukuda \emph{et al.}, editors,
  {\em Mathematical Software — ICMS 2010}, volume 6327 of {\em Lecture Notes
  in Computer Science}, page 35–41. Springer, 2010.

\bibitem{ChudnovskyChudnovsky1990}
D.~V. Chudnovsky and G.~V. Chudnovsky.
\newblock Computer algebra in the service of mathematical physics and number
  theory.
\newblock In D.~V. Chudnovsky and R.~D. Jenks, editors, {\em Computers in
  Mathematics}, volume 125 of {\em Lecture Notes in Pure and Applied
  Mathematics}, page 109–232, Stanford University, 1986. Dekker.

\bibitem{FlajoletSedgewick2009}
P.~Flajolet and R.~Sedgewick.
\newblock {\em Analytic Combinatorics}.
\newblock Cambridge University Press, 2009.

\bibitem{Hille1976}
E.~Hille.
\newblock {\em Ordinary differential equations in the complex domain}.
\newblock Wiley, 1976.
\newblock {D}over reprint, 1997.

\bibitem{Hirschhorn2012}
M.~D. Hirschhorn.
\newblock Estimating the {A}péry numbers.
\newblock {\em The Fibonacci Quarterly}, 50(2):129--131, 2012.

\bibitem{Johansson2013}
F.~Johansson.
\newblock Arb: a {C} library for ball arithmetic.
\newblock {\em ACM Communications in Computer Algebra}, 47(4):166–169, 2013.
\newblock \url{http://fredrikj.net/arb/}

\bibitem{Kauers2013}
M.~Kauers.
\newblock The holonomic toolkit.
\newblock {\em Computer Algebra in Quantum Field Theory: Integration, Summation
  and Special Functions, Texts and Monographs in Symbolic Computation}, 2013.

\bibitem{KauersJaroschekJohansson2015}
M.~Kauers, M.~Jaroschek, and F.~Johansson.
\newblock {O}re polynomials in {S}age.
\newblock In J.~Gutierrez, J.~Schicho, and M.~Weimann, editors, {\em Computer
  Algebra and Polynomials}, page 105–125. Springer, 2015.

\bibitem{Koutschan2013b}
C.~Koutschan.
\newblock Lattice {G}reen's functions of the higher-dimensional face-centered
  cubic lattices.
\newblock {\em Journal of Physics A: Mathematical and Theoretical},
  46(12):125005, 2013.

\bibitem{Lanczos1956}
C.~Lanczos.
\newblock {\em Applied analysis}.
\newblock Prentice-Hall, 1956.

\bibitem{Mallinger1996}
C.~Mallinger.
\newblock Algorithmic manipulation and transformations of univariate holonomic
  functions and sequences.
\newblock Diplomarbeit, RISC-Linz, 1996.

\bibitem{Mezzarobba-numbounds}
M.~Mezzarobba.
\newblock Truncation bounds for {D}-finite series.
\newblock In preparation.

\bibitem{Mezzarobba2010}
M.~Mezzarobba.
\newblock {NumGfun}: a package for numerical and analytic computation with
  {D}-finite functions.
\newblock In S.~M. Watt, editor, {\em {ISSAC} '10}, page 139–146. ACM, 2010.

\bibitem{Mezzarobba2011}
M.~Mezzarobba.
\newblock {\em Autour de l'évaluation numérique des fonctions {D}-finies}.
\newblock Thèse de doctorat, École polytechnique, 2011.

\bibitem{SalvyZimmermann1994}
B.~Salvy and P.~Zimmermann.
\newblock Gfun: A {M}aple package for the manipulation of generating and
  holonomic functions in one variable.
\newblock {\em ACM Transactions on Mathematical Software}, 20(2):163–177,
  1994.

\bibitem{Stanley1999}
R.~P. Stanley.
\newblock {\em Enumerative combinatorics}, volume~2.
\newblock Cambridge University Press, 1999.

\bibitem{Sage}
{The SageMath Developers}.
\newblock {SageMath} mathematics software, 2005–.
\newblock \url{http://www.sagemath.org/}

\bibitem{vdH1999}
J.~van~der Hoeven.
\newblock Fast evaluation of holonomic functions.
\newblock {\em Theoretical Computer Science}, 210(1):199–216, 1999.

\bibitem{vdH2001}
J.~van~der Hoeven.
\newblock Fast evaluation of holonomic functions near and in regular
  singularities.
\newblock {\em Journal of Symbolic Computation}, 31(6):717–743, 2001.

\bibitem{EnckevortStraten2006}
C.~van Enckevort and D.~van Straten.
\newblock Monodromy calculations of fourth order equations of {C}alabi-{Y}au
  type.
\newblock In J.~D. Lewis, S.-T. Yau, and N.~Yui, editors, {\em Mirror Symmetry
  {V}}, volume~38 of {\em AMS/IP Studies in Advanced Mathematics}, page
  539–559. International Press, 2006.

\end{thebibliography}
\end{document}